\documentclass[a4paper,11pt]{article}
\pdfoutput=1 

\usepackage{jinstpub} 

\usepackage{lineno}

\usepackage{subcaption}
\usepackage{siunitx}
\usepackage[super]{nth}

\usepackage{tikz}
\usetikzlibrary{positioning,shapes}

\title{\boldmath Three years of muography at Mount Etna: results and perspectives}

\author[a,g,1]{G.~Gallo,\note{Corresponding author.}}
\author[a,g]{D.~{Lo Presti},}
\author[b]{D.L.~Bonanno,}
\author[c]{G.~Bonanno,}
\author[d,e]{C.~Ferlito,}
\author[a,f]{P.~{La Rocca},}
\author[f]{S.~Reito,}
\author[a,f]{F.~Riggi}
\author[c]{and G.~Romeo}

\affiliation[a]{University of Catania, Department of Physics and Astronomy ``E. Majorana'',\\95123, Catania,Via S. Sofia 64, Italy}
\affiliation[b]{INFN - National Institute for Nuclear Physics - Laboratori Nazionali del Sud,\\95123, Catania,Via S. Sofia 62, Italy}
\affiliation[c]{INAF - National Institute for Astrophysics - Catania Astrophysical Observatory,\\95123, Catania,Via S. Sofia 78, Italy}
\affiliation[d]{University of Catania, Department of Biological, Geophysical and Environmental Sciences,\\95129, Catania,Corso Italia 57, Italy}
\affiliation[e]{INGV - National Institute for Geophysics and Volcanology, Etnean Observatory, Catania section, \\95125, Catania, Piazza Roma 2, Italy}
\affiliation[f]{INFN - National Institute for Nuclear Physics - Catania section,\\95123, Catania, Via S. Sofia 64, Italy}
\affiliation[g]{VMI - International Virtual Muography Institute}

\emailAdd{giuseppe.gallo@unict.it}

\abstract{The Summit Craters system represents the point of maximum expression of the persistent tectonic activity at Mount Etna Volcano. The Muography of Etna Volcano (MEV) Project began in 2016 as a pilot project for the successive installation of a permanent muographic observatory. It aims to demonstrate the detector's capability to observe density anomalies inside the volcanic edifice and monitor their time evolution. The first muon telescope built by the collaboration and installed at the base of the North-East Crater from August 2017 to October 2019 was already able to get significant results. This work describes the characteristics of the muon-telescope and summarizes the principal outcomes obtained, with a quick look at the current status of the project and future developments.}


\arxivnumber{2109.13125} 

\proceeding{22$^{\text{nd}}$ International Workshop on Radiation Imaging Detectors (iWoRiD)\\
  June 27, 2021 to July 1, 2021\\
  Online}

\begin{document}
\maketitle
\flushbottom

\section{Introduction}
\label{sec:intro}

Mount Etna covers about 1,200 square kilometers and is the highest terrestrial volcano on the Eurasian plate, with a height of 3,357 meters a.s.l. \cite{section_etna_2021}. Two fault systems, one NE-SW oriented and the other NNW-SSE, both associated with extensional tectonic activity, intersect at Mount Etna Central Crater (CC). Together with the Northeast (NEC) and Southeast Craters (SEC) and a variable number of fumaroles and vents, they form the ``Summit Craters'' system. Because it is a permanently active volcano, with a surrounding area at high seismic and volcanic hazard, but densely populated, Mount Etna is among the most studied volcanoes in the world. In addition, Sicily's Regional Civil Protection Department and the Etnean Observatory of the National Institute of Geophysics and Volcanology (INGV) monitor it constantly. 

Nevertheless, some questions about Mount Etna and other active volcanoes remain open, such as the ``open-conduit'' model, which describes their plumbing system as an empty cavity filled by the rising magma during the eruptions. Instead, recent observations suggest that the magma dynamically occupies the conduit until it cools down and acts as a plug \cite{ferlito_mount_2018}. The plumbing system can be thought of as divided into three vertical segments: a low-density, high-pressure gas-rich magma fills the lowermost one; above this, there is a high-density basaltic melt, through which gas bubbles can rise from the bottom segment and slowly erode the upper solidified magma; a complex system of surface fractures allowing permanent gas emission constitutes the uppermost part. The vertical extension of each section is continuously varying according to the dynamic equilibrium state until the local tectonics breaks it and an eruption occurs. 

Muography, or muon radiography, has already proved to be the most suitable technique for studying the superficial layers of the volcanic edifice, corresponding to the portion of the structure with a thickness limited to a few kilometers of rock that the most energetic muons can pass through without being stopped. Therefore, its inclusion in a permanent volcano monitoring network could give a fundamental contribution to the understanding of the plumbing system and, thus, to volcanic hazard mitigation \cite{tanaka_radiographic_2014,jourde_improvement_2015,jourde_muon_2016,olah_high-definition_2018,olah_plug_2019}.

\section{The MEV project}

\begin{figure}[hbtp]
	\centering
	
	\begin{subfigure}[t]{.600\linewidth}
		\centering
		\includegraphics[width=\linewidth,trim=400 0 400 700,clip] {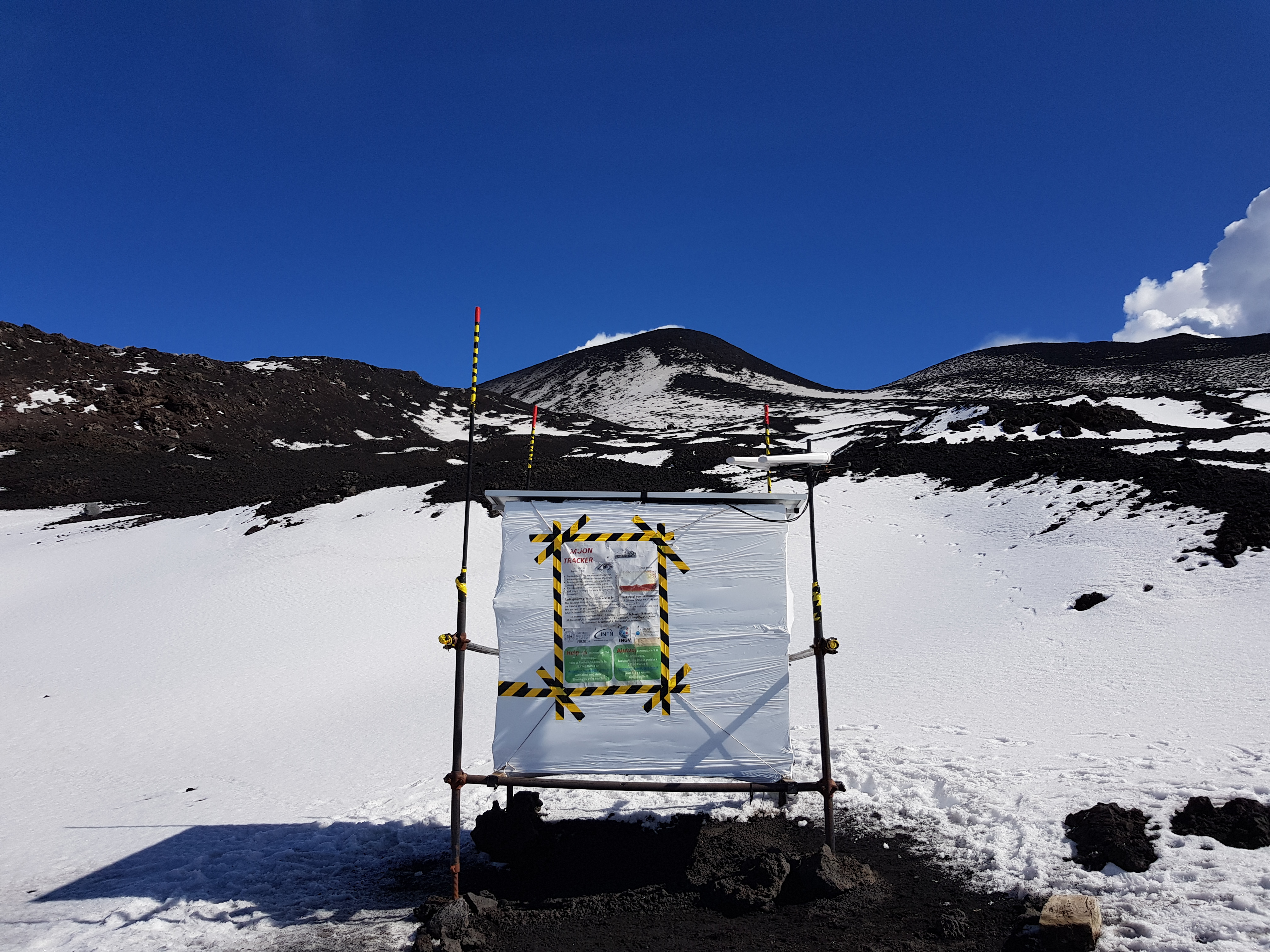}
		\caption{}
		\label{fig:1a}
	\end{subfigure}
	\begin{subfigure}[t]{.385\linewidth}
		\centering
		\includegraphics[width=\linewidth,trim=40 25 85 20,clip] {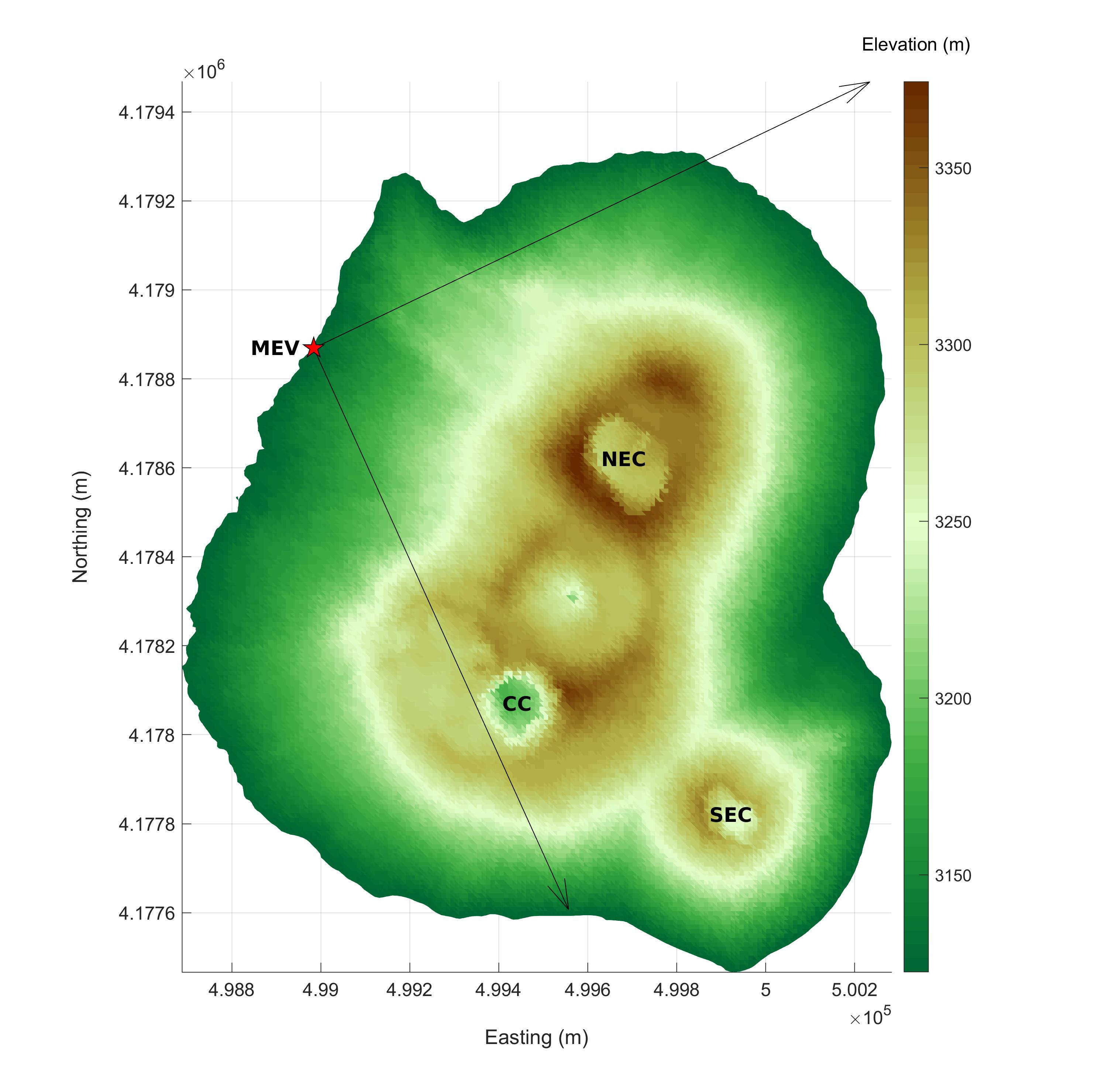}
		\caption{}
		\label{fig:1b}
	\end{subfigure}
	 
	\caption{(a) The telescope placed on the slope of Etna North-East crater at \SI{3100}{\meter} a.s.l. in October 2018. This figure was already published in \cite{lo_presti_muographic_2020}. (b)~Position and field of view of the telescope placed at about \SI{700}{\meter} from the NEC center. The image was elaborated with MATLAB from a Digital Elevation Model updated to December 2015 \cite{ganci_digital_2019}.}
	\label{fig:1}
\end{figure}

In 2016 previously discussed reasons prompted a collaboration of physicists, engineers, and volcanologists to start the Muography of Etna Volcano project to demonstrate the feasibility of a muography experiment in the extreme conditions on the top of Mount Etna. The first muon telescope prototype developed within the project was built at the Department of Physics and Astronomy of the University of Catania (DFA) and transferred to the INGV department of Nicolosi (about \SI{15}{\kilo\meter} far away from Catania) for preliminary tests \cite{lo_presti_mev_2018}. The initial target was the SEC, also to compare new results with previous data \cite{carbone_experiment_2014}, but, at the time of the installation, a lava flow prevented access to it. Then, we moved to the NEC for the test and placed the telescope in this area, as shown in figure~\ref{fig:1}.

\subsection{The muon telescope}
A muon telescope is essentially a stack of at least two position-sensitive detectors arranged to get the impact points of the particles crossing it and reconstruct their trajectories. Therefore, it allows measuring the particle flux for each direction within its field of view to estimate the attenuation due to the target object. For this reason, the telescope must be placed and oriented so that the subject of interest is the unique cause of muon flux attenuation. We designed our telescope to have its axis, perpendicular to the detection modules, horizontally aligned to measure at the same time the muon flux from the side of NEC and the one from the opposite direction without any source of mitigation referred to in the following as ``open-sky'' flux.

The design of the first muon telescope built within the MEV project includes all the requirements for an out-of-lab application: a container box ensures ruggedness and water-tightness; solar panels and battery pack to be independent of the power network; just \SI{25}{\watt} of power needed; remote control and data transfer over an internet connection.
The telescope has three position-sensitive modules with one square meter area and a modular aluminum frame that allows the alignment. The detector and all the stuff required weight about \SI{300}{\kilogram}. Thus, it requires a four-wheel-drive truck with a mechanical arm for transporting and installing it already mounted. In principle, this design allows you to carry the pieces one at a time and assemble the telescope in place. See ref.~\cite{lo_presti_mev_2018} for a more detailed description of the telescope.

\subsection{The position-sensitive modules}
\begin{figure}
	\centering
	
	\noindent
	\begin{minipage}[c]{0.50\textwidth}
		\begin{subfigure}[t]{\linewidth}
			\centering
			\includegraphics[width=\linewidth,trim=0 0 0 0,clip] {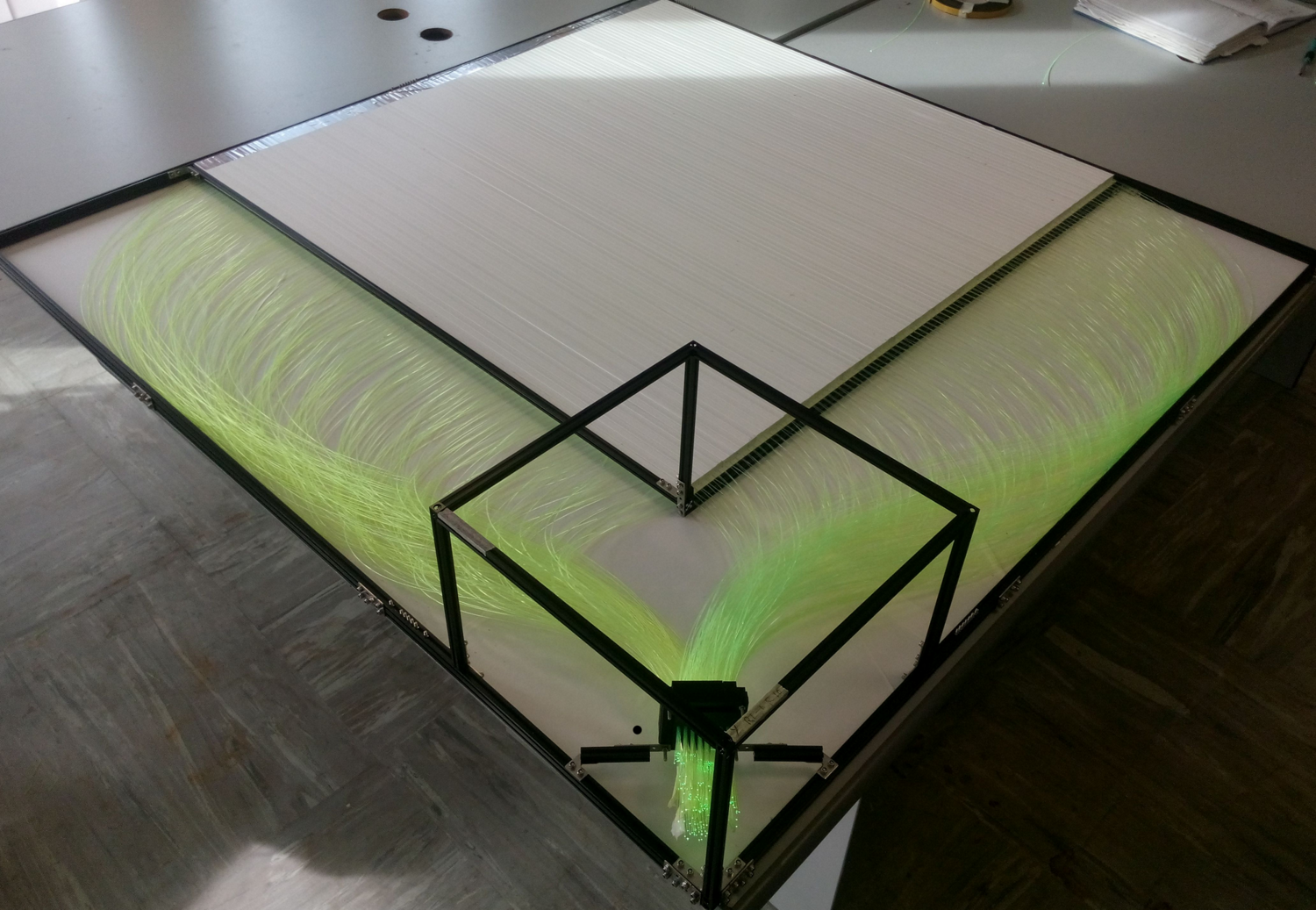}
			\caption{}
			\label{fig:4a}
		\end{subfigure}
		\begin{subfigure}[t]{\linewidth}
			\centering
			\includegraphics[width=\linewidth,trim=0 250 0 100,clip] {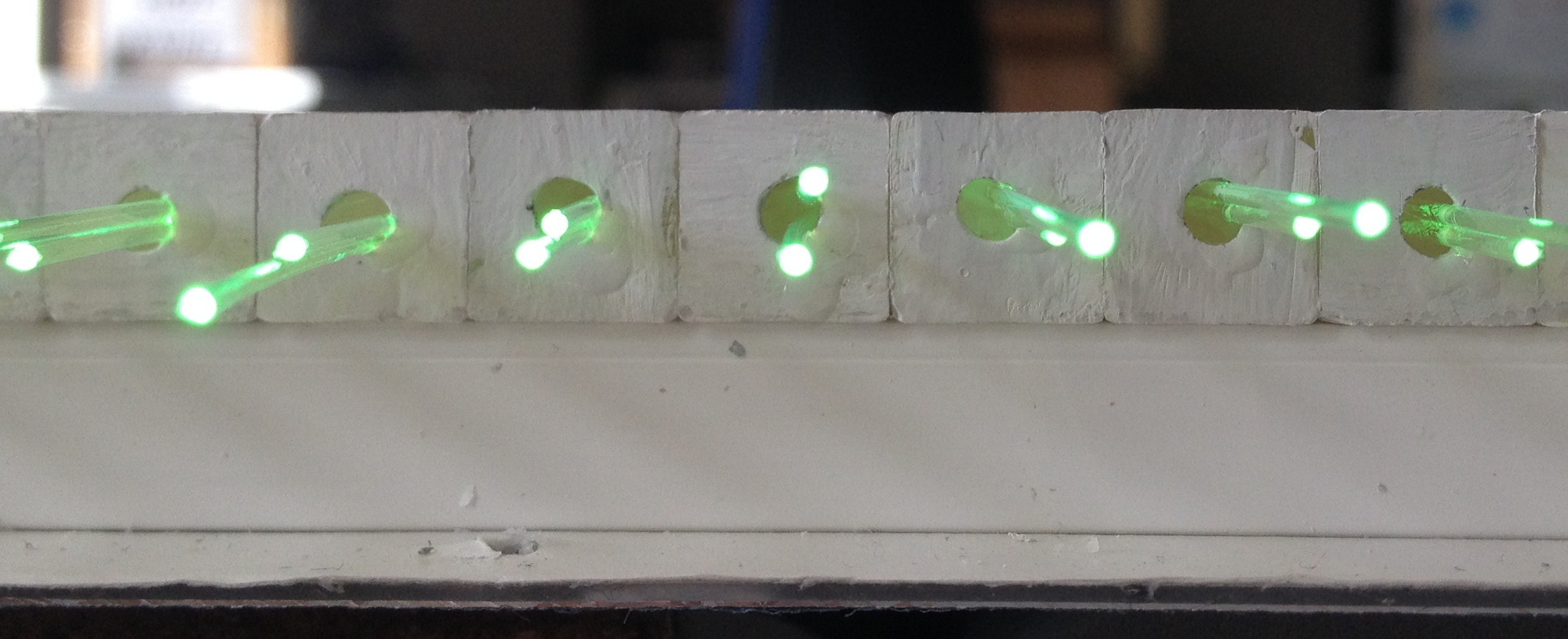}
			\caption{}
			\label{fig:4b}
		\end{subfigure}
	\end{minipage}%
	\quad
	\begin{minipage}[c]{0.30\textwidth}
		\begin{subfigure}[t]{\linewidth}
			\centering
			\includegraphics[width=\linewidth,trim=20 1 20 40,clip] {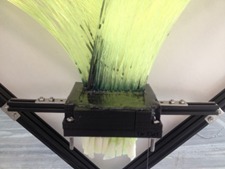}
			\caption{}
			\label{fig:4c}
		\end{subfigure}
		\begin{subfigure}[t]{\linewidth}
			\centering
			\includegraphics[width=\linewidth] {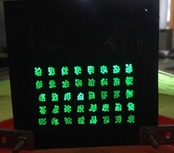}
			\caption{}
			\label{fig:4d}
		\end{subfigure}
	\end{minipage}
	
	\caption{Details of a telescope position-sensitive module assembly. (a)~Black painted aluminum frame enclosing the two layers of extruded scintillating bars. (b)~Detail of the two WLS fibers embedded in each bar. (c)~Top view of the 3D printed plastic frame for WLS-PMT coupling. (d)~Bottow view of the frame after cutting and lapping of the WLS.}
	\label{fig:4}
\end{figure}

Each position-sensitive module consists of two orthogonal layers of 99 scintillating bars with nominal size $1\times1\times100$~\si{\centi\meter}. They have a central hole in which two rounded 1-millimeter Wavelength Shifting fibers (WLS) are embedded to transport the scintillation light to a Multi Anode PMT. A 3D printed plastic frame allows an easy coupling between the WLS bundle and the PMT. Figure~\ref{fig:4} shows some pictures taken during the assembly of a module. The distance between the outer modules of the telescope, which is \SI{97}{\centi\meter}, and the number and pitch of pixels for each module define its geometric characteristics. The angular aperture is $\pm45$~degrees; the resolution is about \SI{10}{\milli\radian}. In the muographic images shown in this paper a pixel corresponds to a $7\times7$~\si{\square\meter} target area because of the geometric characteristics of the telescope just described and its position about \SI{700}{\meter} away from the central axis of the NEC.

\subsection{Front-End, data processing \& control unit electronics}
Each module has its single Front-End (FE) board equipped with a 64 channels Multi-Anode Photomultiplier (MAPMT, mod. Hamamatsu H8500)  to read out the scintillation light signal. The MAROC3 chip is the core element of the FE architecture. It is an ASIC designed for reading MAPMT signals. It performs the pre-amplification and shaping of the analog signal and the analog-to-digital conversion, giving a time-over threshold response for each channel.

Counting both WLS for each scintillating bar, there are 396 ($4\times99$) optical channels for each module, but the sensor has 64 ones only. The coupling is made possible by a proper routing of the WLS. It allows minimizing the number of corresponding FE channels by a factor of 10 ($1/\sqrt{99}$). Thus, the number of optical channels per module is equal to 40 (see figure~\ref{fig:4d}). The reduction mechanism is quite simple. The WLS fibers inside each bar have two different labels, \textit{Group} and \textit{Strip}, respectively. If $i$ is the bar progressive index from 1 to 99, the corresponding \textit{Group} $(G)$ and \textit{Strip} $(S)$ can be calculated as:
\begin{align*}
G &= 1 + (i-1)\ \mathbf{div}\ n \\
S &= (i-1)\ \mathbf{rem}\ n
\end{align*}
using the integer division ($\mathbf{div}$) and reminder ($\mathbf{rem}$) operators. The integer $n$ is equal to the number of bars belonging to the same \textit{Group}-set. In this case $n$ is equal to $10$, resulting in ten \textit{Group-} and as many \textit{Strip-}sets, which are 20 channels per layer. The two WLS of each bar are differently routed according to the \textit{Group} and \textit{Strip} labels. The bar hit by the particle is unambiguously reconstructed as $i = (G-1)\times n + S$.

All sensitive modules share the unique data processing and control board. Its main component is the System-On-Module (SOM) by National Instruments, an embedded computer with a processor, a custom Linux Real-Time operating system, and a programmable FPGA. A dedicated UI allows controlling all the parameters of the system and managing the acquisition. The board holds an SD card to save data locally, but the system has access to the internet using a 4G wireless router and periodically transfers the data to the network storage hosted at the DFA. 

All the electronic boards are custom designed to remove the unnecessary components found in MAROC3 and SOM's evaluation boards, reducing the power consumption. See ref.~\cite{lo_presti_mev_2018} for more details on the electronics.

\section{State of the project and results}
\begin{figure}[hbtp]
	\centering
	
	\begin{subfigure}[t]{.500\linewidth}
		\centering
		\includegraphics[width=\linewidth] {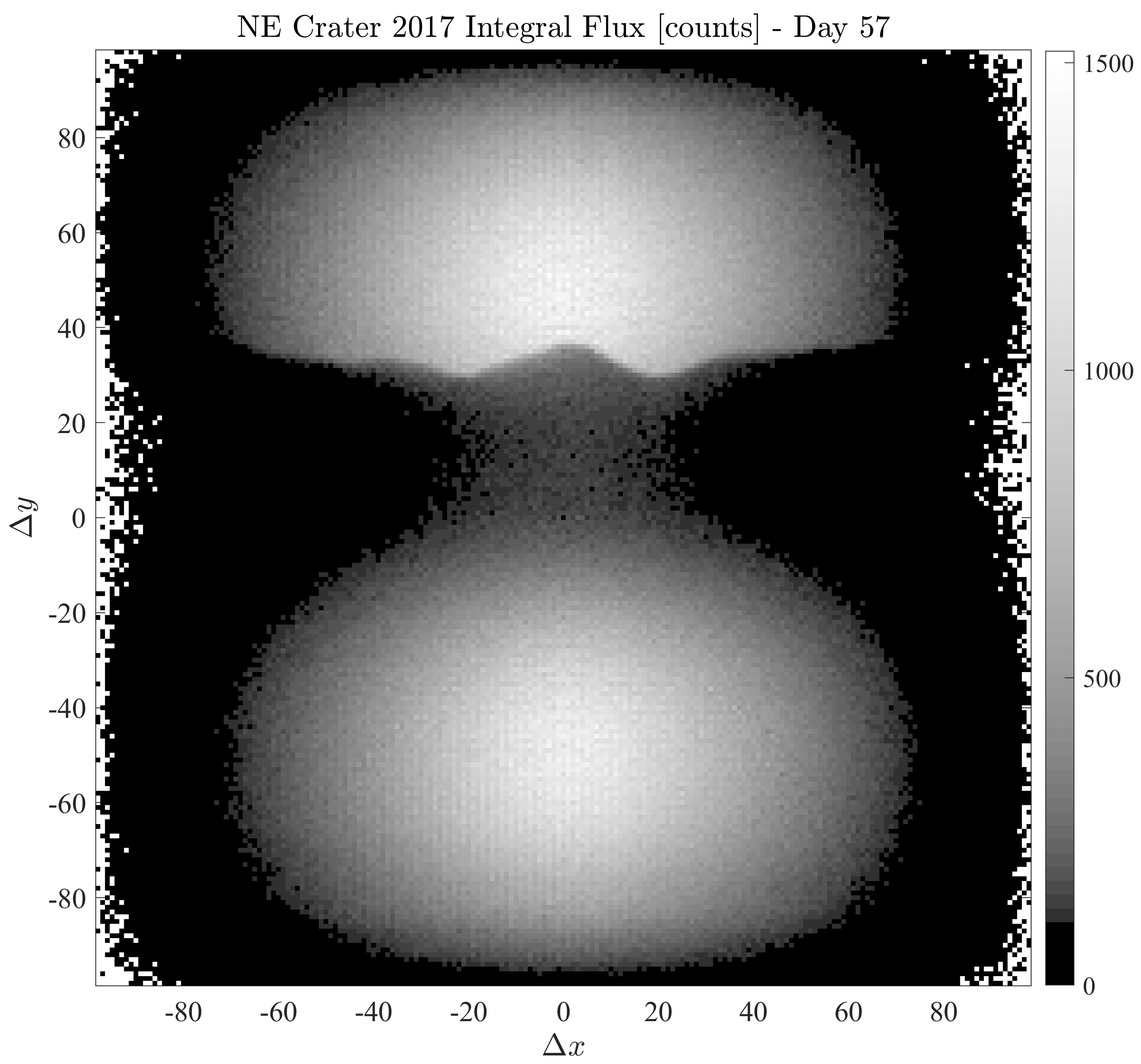}
		\caption{}
		\label{fig:2a}
	\end{subfigure}
	\begin{subfigure}[t]{.485\linewidth}
		\centering
		\includegraphics[width=\linewidth] {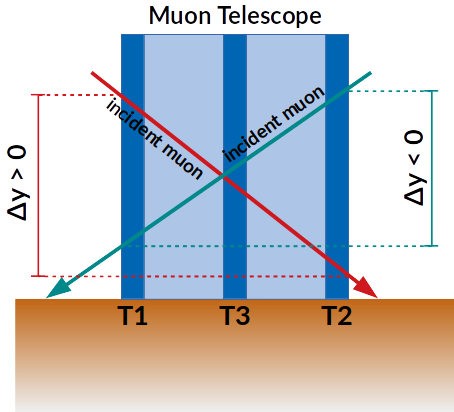}
		\caption{}
		\label{fig:2b}
	\end{subfigure}
	
	\caption{(a) Muon integral flux acquired in 2017, starting from August \nth{1} till the telescope shut-down due to the snow coverage on the solar panels. (b) Sketch of telescope detection modules T1, T2, and T3, with the incoming direction distinction by $\Delta y$.}
	\label{fig:2}
\end{figure}

Figure~\ref{fig:2a} shows an example of muon imaging of the NEC in terms of integral flux measured. In such an image, each ($\Delta x$, $\Delta y$) bin corresponds to a particular direction. $\Delta x$ and $\Delta y$ are the differences in the number of pixels between the entry and exit points of the particles on the two outermost detection modules. The sign of $\Delta y$ distinguishes muons coming from the two opposite sides of the telescope, as sketched in figure~\ref{fig:2b}.

The permanence of the telescope at high altitude from August 2017 until October 2019 has allowed demonstrating the feasibility of permanent monitoring of the Etna Summit Craters system, the primary objective of the measurement campaign, and the muographic results obtained are significant. During winter, the snow buried the solar panels that powered the detector; it was necessary to go to the installation site to restart the acquisition in spring. Thus, it was only possible to acquire during the summer for 57 days in 2017, 95 days in 2018, and 51 days in 2019, respectively. However, this issue can easily be overcome by a better installation of solar panels and by constantly removing snow before it becomes too thick.

The following muographic images show the results in terms of $R$, the daily average of the ratio between the flux of particle surviving after traversing the crater and the one before it. Because we have no direct access to this last quantity, the muon flux of the open-sky, measured from the opposite side of the telescope, replaces it. See refs.~\cite{gallo_improvements_2019,riggi_investigation_2020} for the complete data analysis procedure and the evaluation of the East-West effect at the measurement site. 

\begin{figure}[hbtp]
	\centering
	\hspace{-6pt}
	\begin{subfigure}[c]{.670\linewidth}
		\centering	
		\begin{tikzpicture}
			\node(a){\includegraphics[width=\linewidth,trim=0 0 5 0,clip] {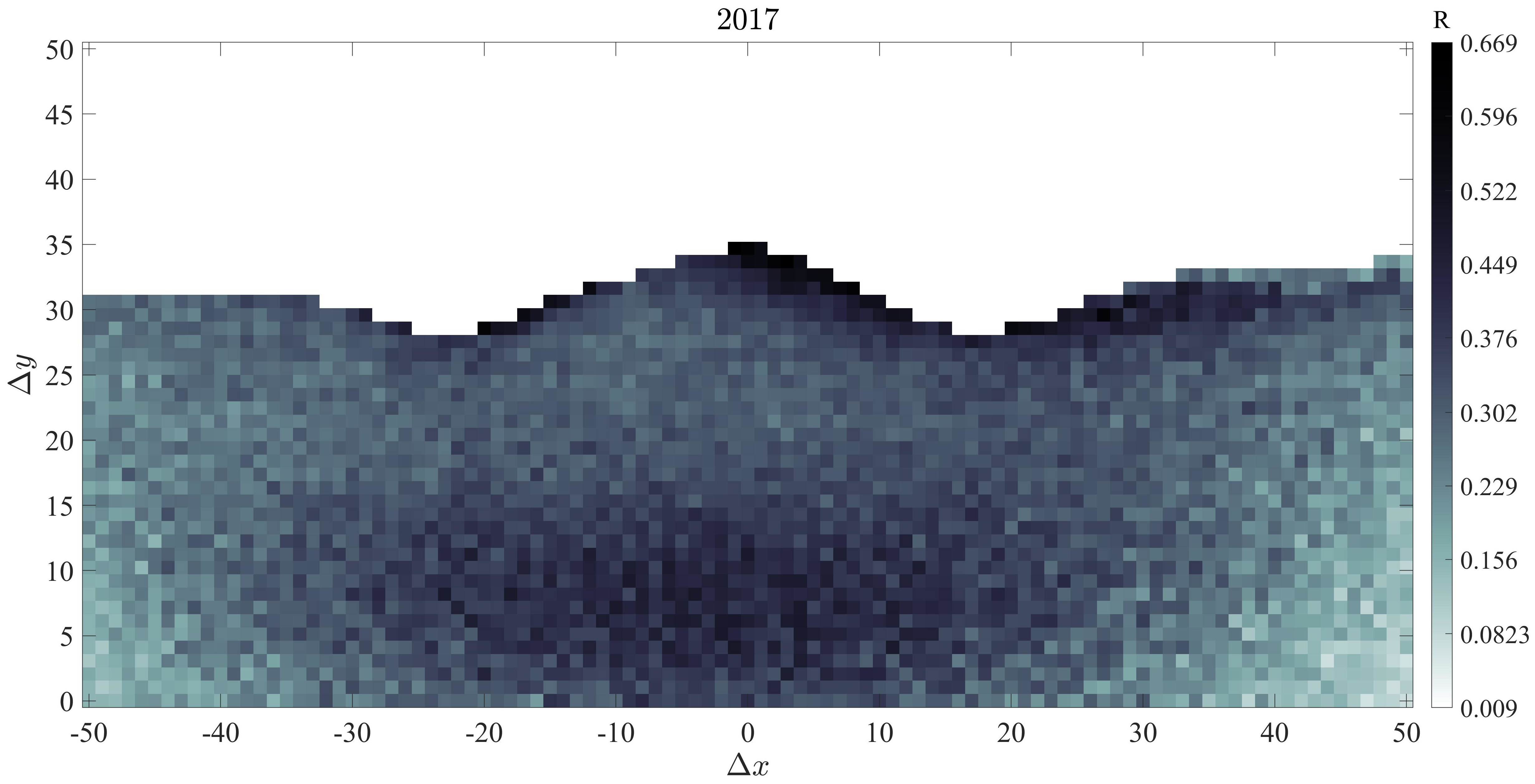}};
			\node at(a.center)[draw, red,line width=3pt,ellipse, minimum width=110pt, minimum height=40pt,rotate=0,yshift=-40pt,xshift=-10pt]{};
		\end{tikzpicture} 
	\quad
	\end{subfigure}
	\hspace{3pt}
	\begin{subfigure}[c]{.300\linewidth}
		\centering
		\includegraphics[width=\linewidth,trim=0 0 0 0,clip] {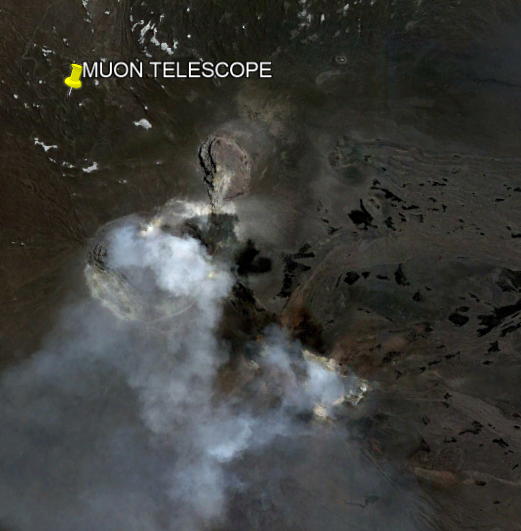}
		\quad
	\end{subfigure}

	\hspace{-6pt}
	\begin{subfigure}[c]{.670\linewidth}
		\centering	
		\begin{tikzpicture}
			\node(a){\includegraphics[width=\linewidth,trim=0 0 5 0,clip] {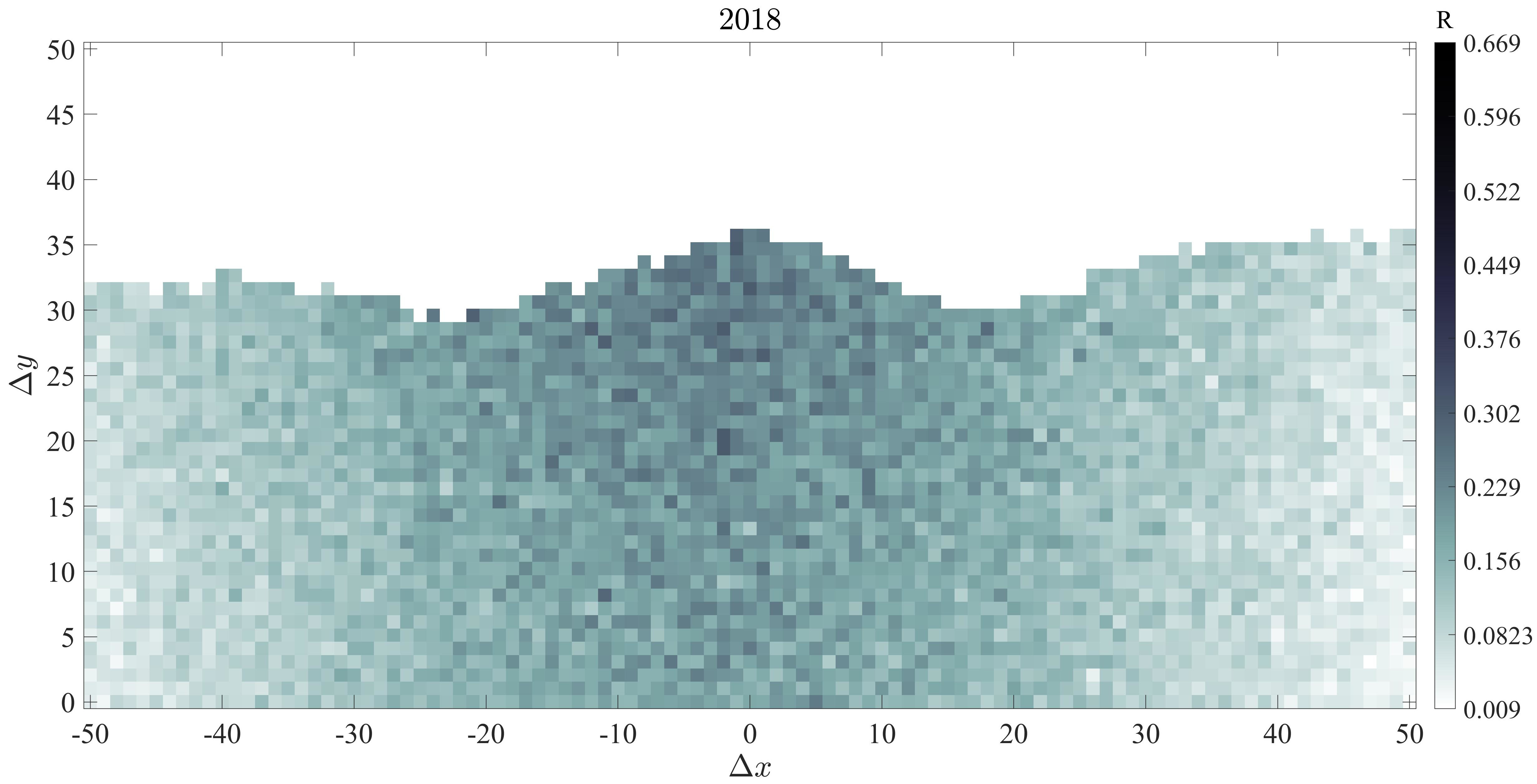}};
			\node at(a.center)[draw, dotted, red,line width=2pt,ellipse, minimum width=80pt, minimum height=40pt,rotate=60,yshift=-45pt,xshift=20pt]{};
		\end{tikzpicture} 
		\quad
	\end{subfigure}
	\hspace{3pt}
	\begin{subfigure}[c]{.300\linewidth}
		\centering
		\includegraphics[width=\linewidth,trim=0 0 0 0,clip] {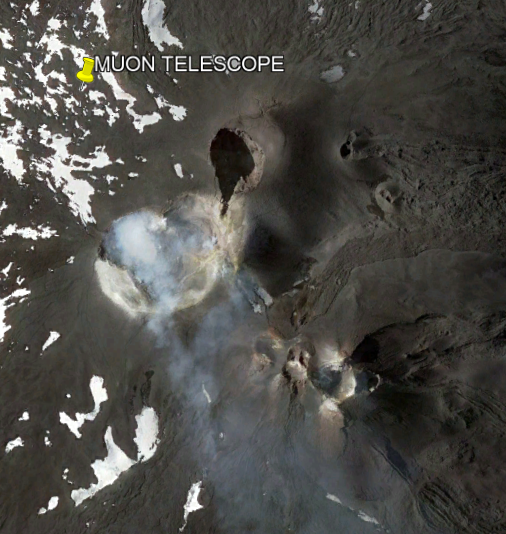}
		\quad
	\end{subfigure}

	\hspace{-6pt}
	\begin{subfigure}[c]{.670\linewidth}
		\centering	
		\begin{tikzpicture}
			\node(a){\includegraphics[width=\linewidth,trim=0 0 5 0,clip] {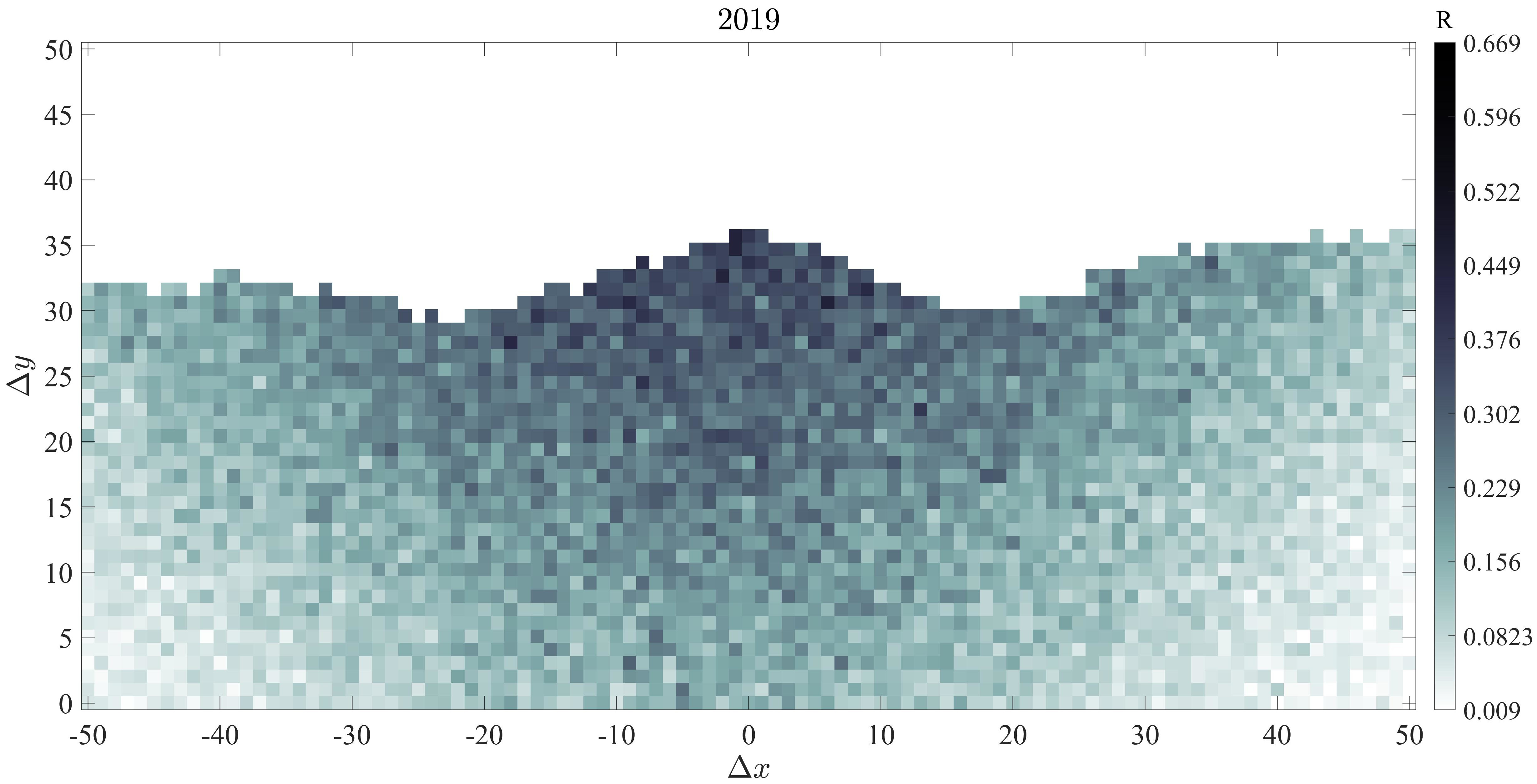}};
			\node at(a.center)[draw, dotted, red,line width=2pt,ellipse, minimum width=80pt, minimum height=40pt,rotate=60,yshift=-45pt,xshift=20pt]{};
		\end{tikzpicture} 
	\end{subfigure}
	\hspace{3pt}
	\begin{subfigure}[c]{.300\linewidth}
		\centering
		\includegraphics[width=\linewidth,trim=0 70 0 30,clip] {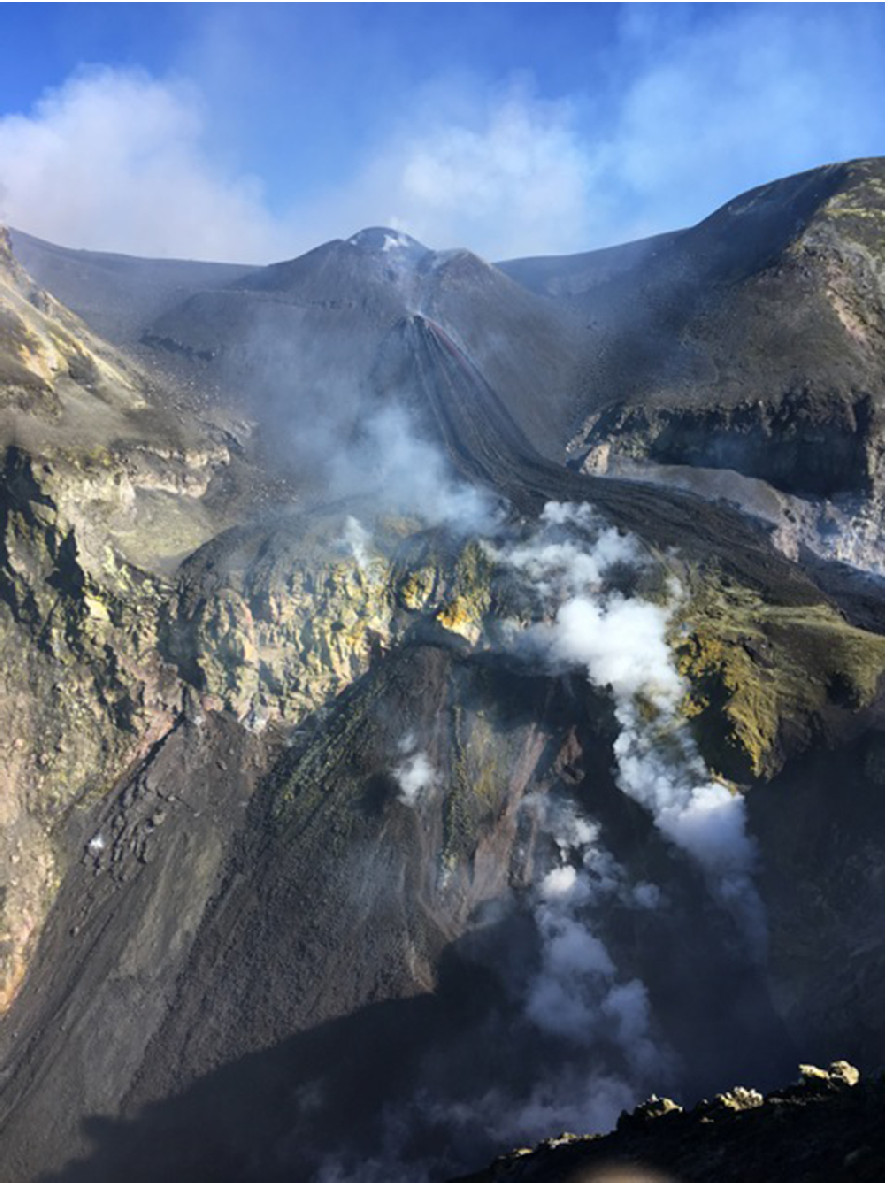}
	\end{subfigure}
		
	\caption{Comparison between 2017, 2018, and 2019 muographic images on the left panels, with the corresponding satellite views (Google Earth screen captures) for 2017 and 2018. The bottom right corner picture was taken in February, 2020 and shows a lava pit opened on the inner flank of the CC.}
	\label{fig:3}
\end{figure}
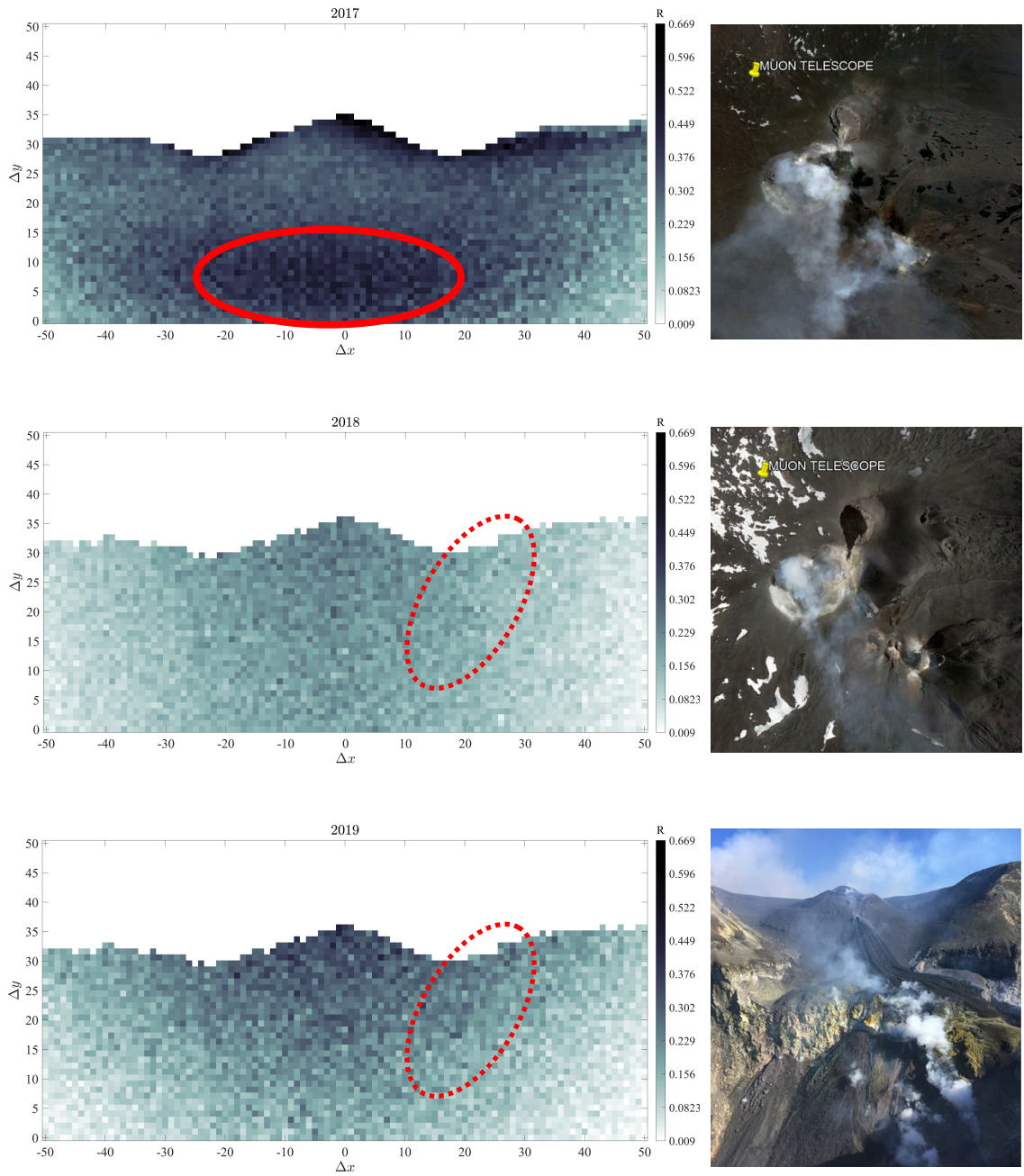

One of the main volcanological findings by this research is the evidence of a cavity under the NEC roof, detected months before its collapse at the end of 2017. The comparison between 2017 and 2018 $R$ maps in figure~\ref{fig:3} gives an immediate hint of this anomaly. In 2017 muography, the region of $R$ abnormally higher than its surroundings (inside the red ellipse) is the evidence of an unexpectedly high muon flux caused by a lower density material along the muon path or an empty volume. In 2018 muography, there is a V-shaped region of high $R$, extending from the surface to the previous location of the cavity and, in fact, from the satellite views (figure~\ref{fig:3}, right panels) and the visual evidence on-site, we can see the NEC without its roof after the collapse. See ref.~\cite{lo_presti_muographic_2020} for a comprehensive volcanological discussion of these results.

Another interesting phenomenon suggested by etnean models and confirmed by muography and successive visual evidence is the fractures migration from the North-East to the Central crater. In both the 2018 and 2019 muography, the dashed red ellipses highlight some higher $R$ spots along the direction connecting the NEC to the CC.  The opening of a new lava pit on the inner flank of the CC in February 2020, right along this line, further confirmed the consistency of these observations (see figure~\ref{fig:3}, bottom right panel).

\subsection{The time-of-flight module}
The purpose of this measurement is the correct discrimination of near-horizontal tracks. This very tiny flux suffers the noise by upward-going muons. They are particles that seem to come from below the horizontal plane before passing through the telescope so that they perfectly mimic particles that come downwards in the opposite direction and produce a wrong increase of the flux \cite{jourde_experimental_2013,gomez_forward_2017}.

The TOF module consists of an additional electronic board (TDC-GPX2 evaluation board) which receives the OR signal output of the MAROC3 chip in each board after a delay line, a monostable and a digital level translator. Counting the time jitter of all elements involved in the TOF measurement chain, the time resolution is about \SI{1}{\nano\second}. 

\begin{figure}[hbtp]
	\centering
	
	\begin{subfigure}[t]{.710\linewidth}
		\centering
		\includegraphics[width=\linewidth,trim=0 0 5 0,clip] {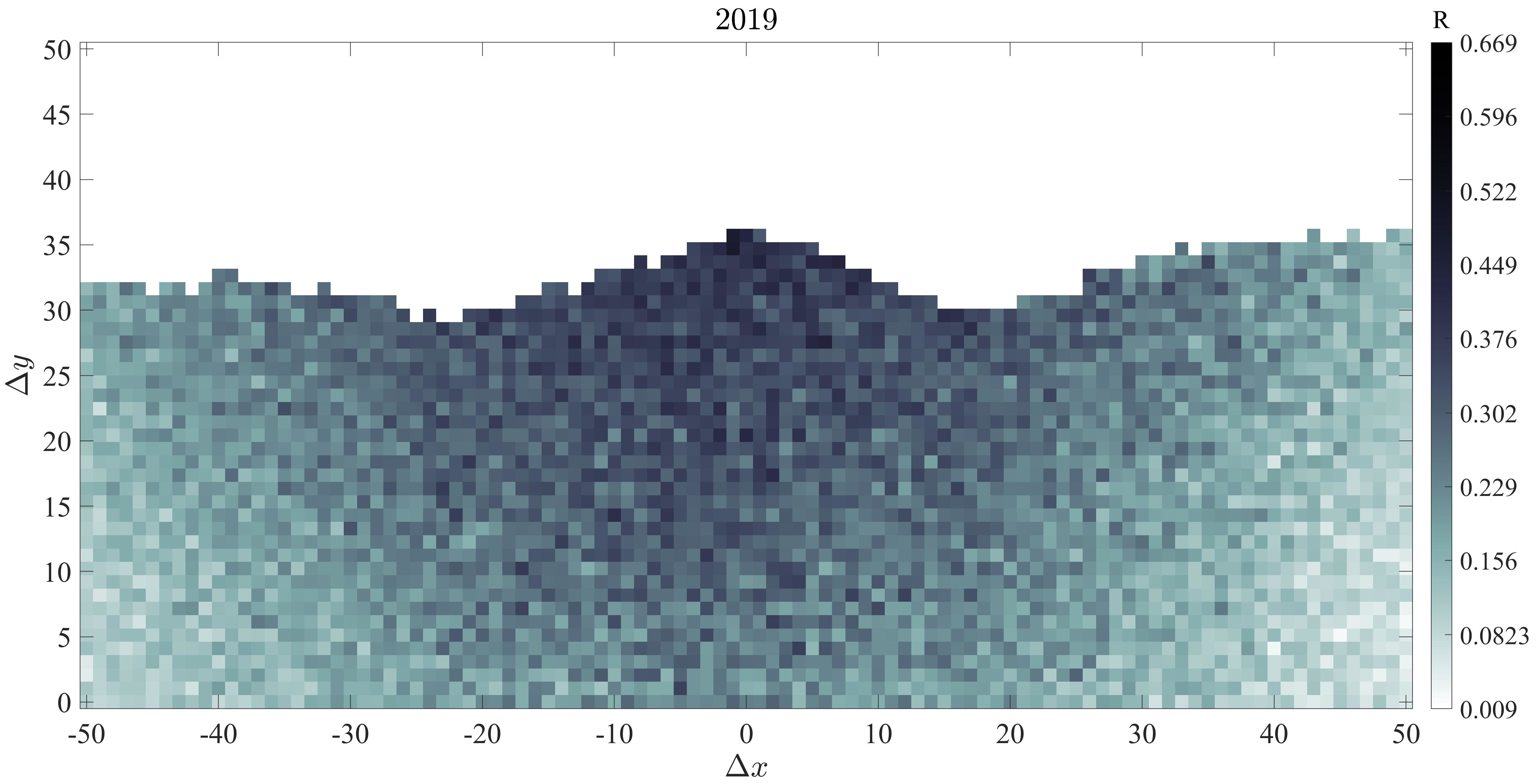}
		\caption{}
		\label{fig:5a}
		\quad
	\end{subfigure}
	\begin{subfigure}[t]{.710\linewidth}
		\centering
		\includegraphics[width=\linewidth,trim=0 0 5 0,clip] {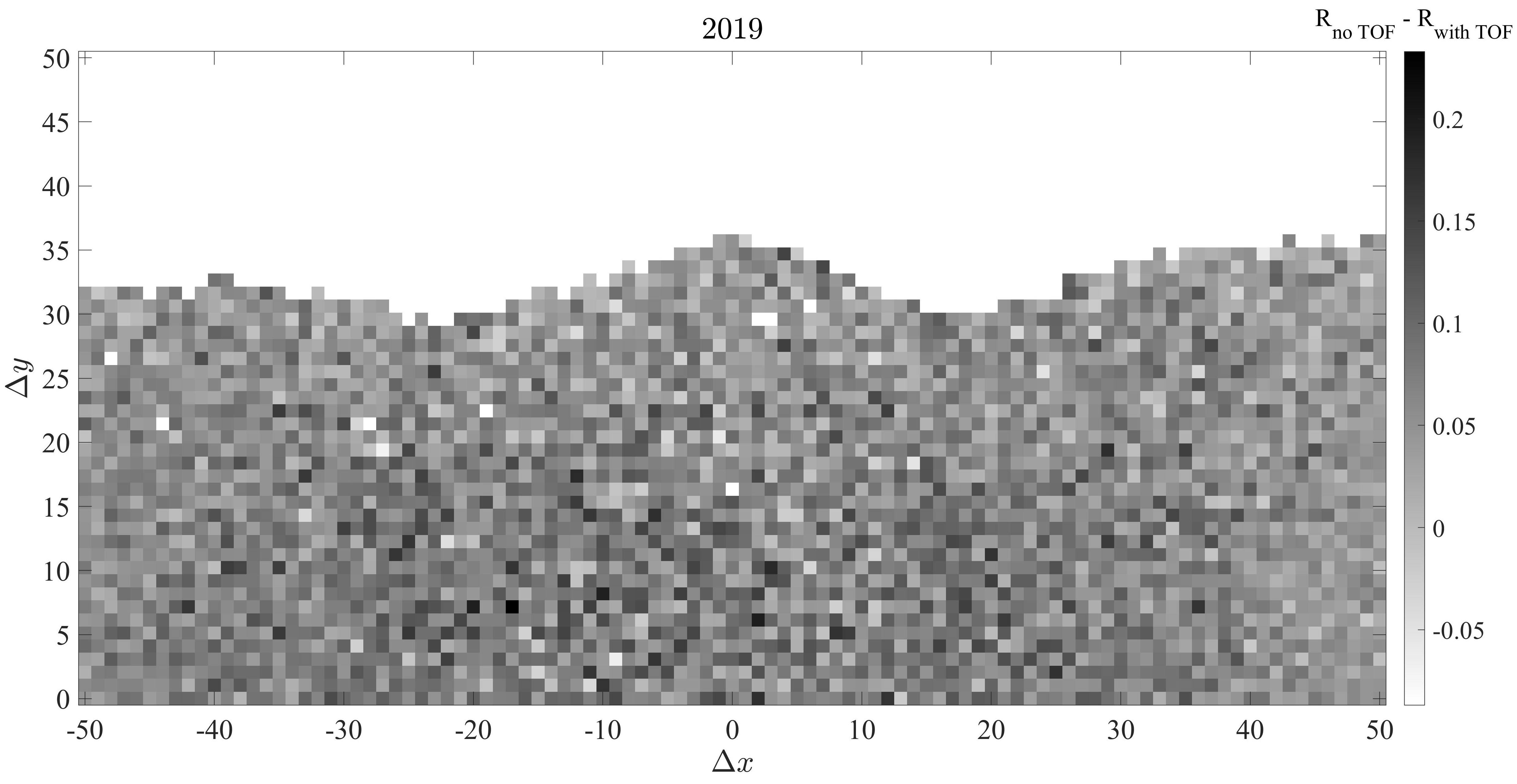}
		\caption{}
		\label{fig:5d}
		\quad
	\end{subfigure}
	\begin{subfigure}[t]{.485\linewidth}
		\centering
		\includegraphics[width=\linewidth] {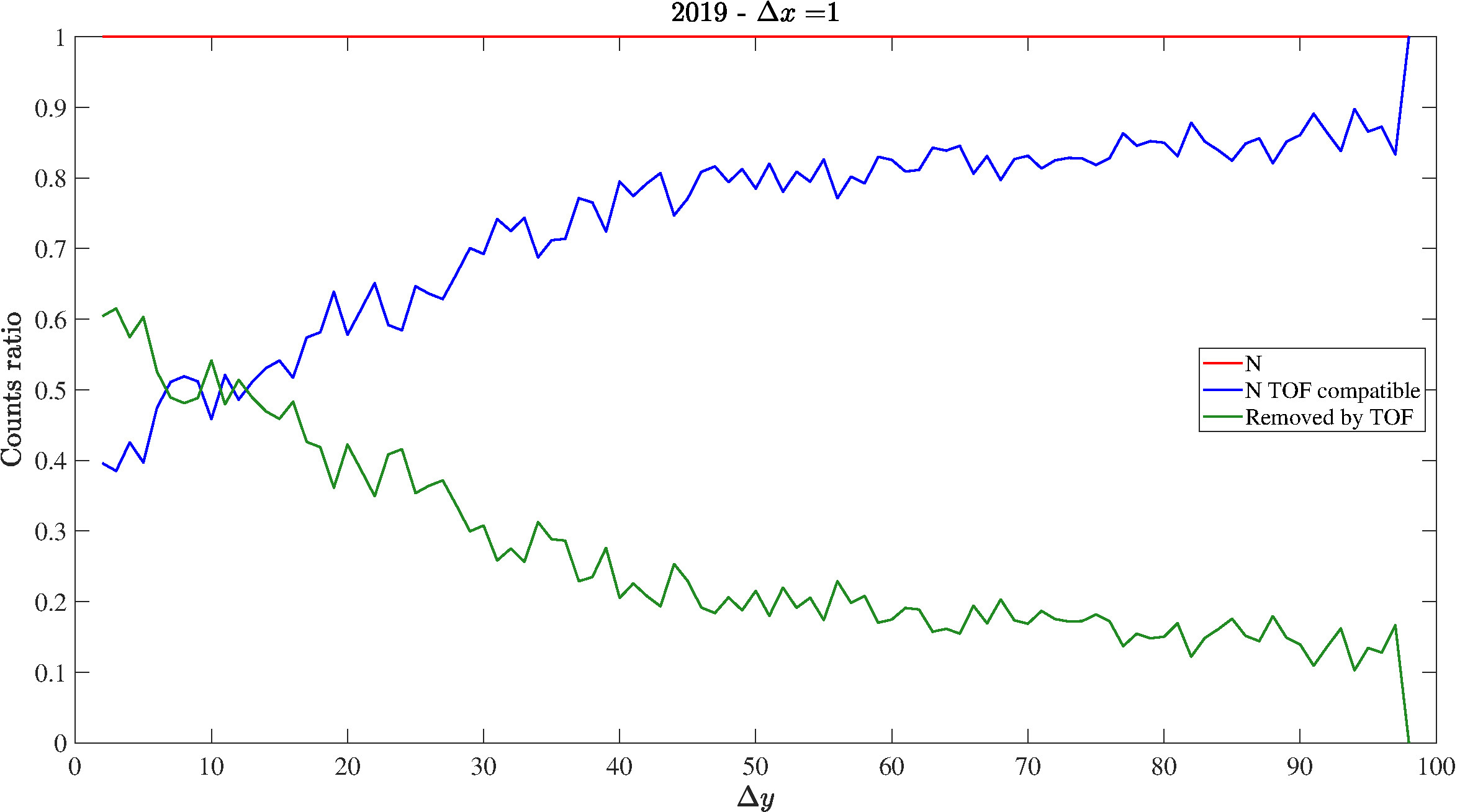}
		\caption{}
		\label{fig:5b}
	\end{subfigure}
	\hfill
	\begin{subfigure}[t]{.485\linewidth}
		\centering
		\includegraphics[width=\linewidth] {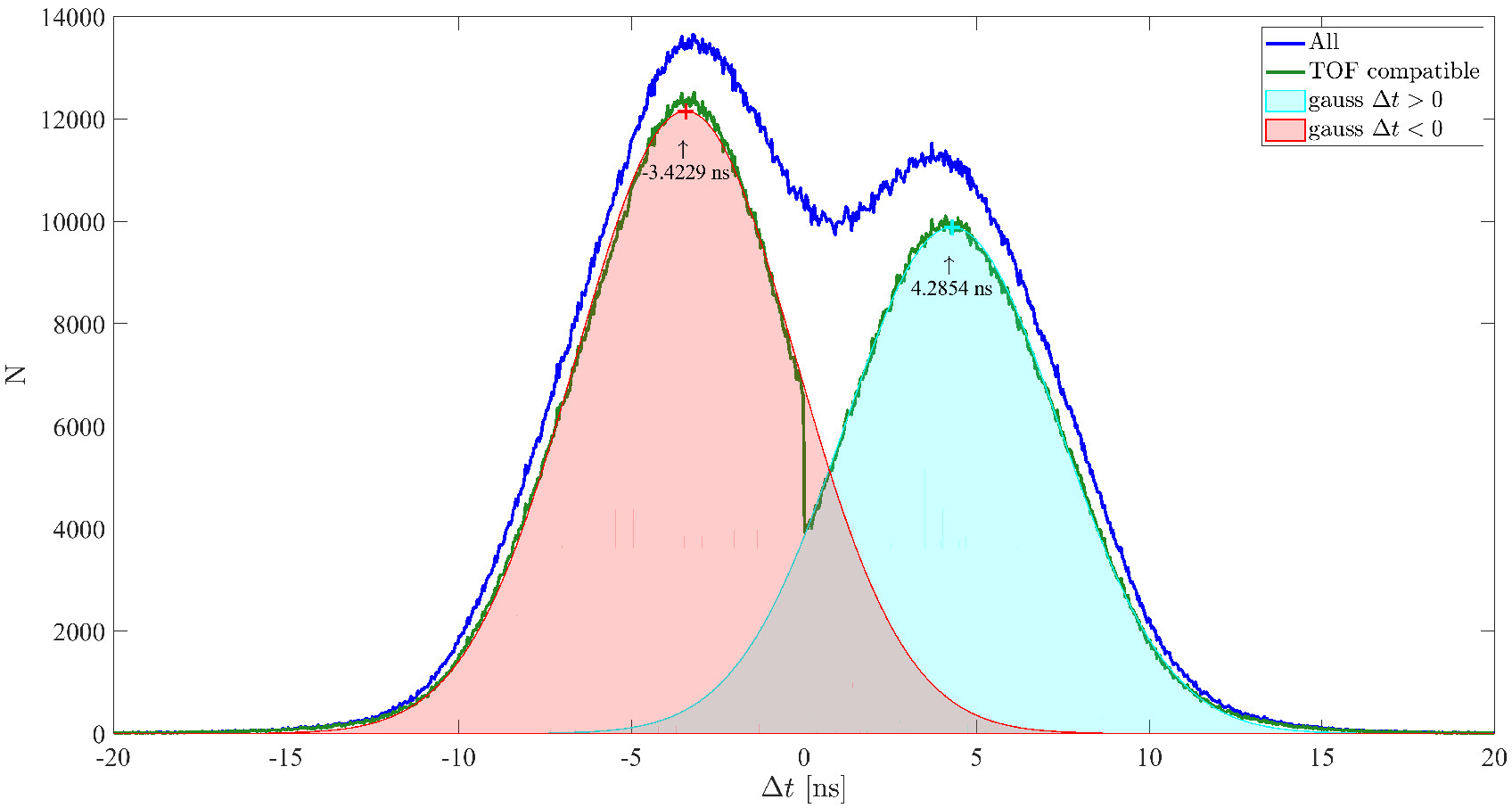}
		\caption{}
		\label{fig:5c}
	\end{subfigure}
	
	\caption{(a)~2019 NEC muography without TOF filtering. (b)~Difference between $R$ values in 2019 muographs without and with TOF filter. (c)~Red: normalized counts of  particle flux incoming from the target side. Blue: normalized number of events for which the incoming direction is uniquely determined by both $\Delta y$ and TOF measure. Green: normalized number of rejected events. (d)~Overall (blue) and filtered (green) $\Delta t$ distributions for data acquired during 2019. Red and cyan shaded areas represent two Gaussian distributions used to fit independently filtered positive and negative $\Delta t$ histograms. Values indicated by crosses correspond to the peaks' center position.}
	\label{fig:5}
\end{figure}
The TOF measurement provides a further possibility to discriminate the direction of arrival of the particles, in addition to the geometric criterion of $\Delta y$. The comparison between the 2019 muographic images with the TOF filter applied (figure~\ref{fig:3}) to the one without (figure~\ref{fig:5a}, also shown as the differential map in figure~\ref{fig:5b}), clearly shows its effect. As one would expect, the percentage of rejected events increases approaching the horizontal or decreasing $\Delta y$ (see the green line in figure~\ref{fig:5b}). However, with this resolution and the external modules placed about 1 meter apart it is impossible to completely separate the two distributions of particles entering the detector from the front or the rear side (figure~\ref{fig:5c}).

\section{Conclusion}

Muographic results here discussed sustain the models of etnean dynamics proposed by the volcanologists of the collaboration and vice-versa; visual evidence confirmed both. The technology and the ingenious solution adopted demonstrated an optimal fit for this application. The international muographers community appreciated the MEV project such that the Universities of Catania and Tokyo and the Wigner Research Center signed an international research agreement in September 2019. The plans for this activity include the TOF measurement improvement by increasing the telescope length up to \SI{2.5}{\meter}, and the reinforcement of the collaboration with the Italian Institute for Volcanology to integrate the muographic monitoring with the existing sensors network.




\bibliographystyle{JHEP}
\bibliography{Gallo_iWoRiD_proceeding.bib}


%
%
%



\end{document}